\documentclass[letterpaper, 10 pt, conference]{ieeeconf} 

\IEEEoverridecommandlockouts  
\usepackage{amsmath,amsfonts}
\usepackage{algorithmic, algorithm}
\usepackage{color}
\usepackage{array}
\usepackage{fixltx2e}
\usepackage{url}
\usepackage{graphicx,graphics}
\usepackage{subfigure}
\usepackage{caption}
\usepackage[
style = ieee,
sorting=none,
backend=bibtex]{biblatex}
\begin{document}

\title{\LARGE{ \bf Exact Consistency Tests for Gaussian Mixture Filters \\ using Normalized Deviation Squared Statistics}}
%
%
%

\author{Nisar Ahmed$^*$,
\thanks{$^*$ Smead Aerospace Engineering Sciences, University of Colorado Boulder, Boulder, CO, 80309.}
Luke Burks$^\dagger$, Kailah Cabral$^\dagger$, Alyssa Bekai Rose$^\dagger$
\thanks{$^\dagger$ Aurora Flight Sciences, a Boeing Company, Cambridge, MA, 02142.}
}
\maketitle

\begin{abstract}
We consider the problem of evaluating dynamic consistency in discrete time probabilistic filters that approximate stochastic system state densities with Gaussian mixtures. 
Dynamic consistency means that the estimated probability distributions 
correctly describe the actual 
uncertainties. As such, the problem of consistency testing naturally arises in applications with regards to estimator tuning and validation. However, due to the general complexity of the density functions involved, straightforward approaches for consistency testing of mixture-based estimators  
have remained challenging to define and implement. This paper derives a new exact result for Gaussian mixture consistency testing within the framework of normalized deviation squared (NDS) statistics. 
It is shown that NDS test statistics for generic multivariate Gaussian mixture models exactly follow mixtures of generalized chi-square distributions, for which efficient computational tools are available. The accuracy and utility of the resulting consistency tests are numerically demonstrated on static and dynamic mixture estimation examples. 
\end{abstract}


%
\IEEEpeerreviewmaketitle

\section{Introduction} \label{sec:intro}
The problem of discrete-time state estimation with 
non-Gaussian uncertainties arises in many settings. It can be addressed via recursive Bayesian filtering algorithms, which leverage prior models of system dynamics and measurement processes to construct posterior probability density functions (pdfs) of the states, conditioned on available observations. Since posterior pdfs cannot in general be exactly obtained, approximations are needed to trade off losses in representational fidelity of uncertainty for gains in computational tractability. This includes linear estimators that attempt to approximate the state mean and error covariance 
(e.g. extended and unscented Kalman filters), as well as those which approximate the full state posterior pdf, such as the Gaussian mixture filter and particle filter algorithms.  

In addition to their different pros and cons from an implementation standpoint, different approximations incur different information losses relative to the exact posterior distribution and associated state statistics. 
Moreover, each approach relies on different tuning parameters to permit flexibility across a range of different problem settings. Therefore a major goal of estimator design is to optimally balance the acceptability of information loss relative to expected accuracy and precision measures, quantified by estimation error bias, variance, mean square error, etc. 
To this end, statistical methods for evaluating \emph{dynamic consistency} serve as powerful tools for estimator tuning and provide rigorous necessary conditions for formal estimator validation. 

Dynamic consistency indicates that a 
filter correctly describes 
actual 
uncertainties with respect to estimation errors and observed data. As such, estimators can be evaluated for consistency using ground truth model simulations and/or real observation data logs 
by comparing a filter's model of its error uncertainties to the true underlying distribution/statistics. 
However, the ability to make such formal statistical comparisons in practice depends heavily on the application and type of estimator used. In particular, while statistical consistency tests are well established for linear(ized) filters with Gaussian priors, process noise, and measurement noise distributions \cite{barshalom2001estimation}, these tests are invalid for mixture filters in settings with highly non-Gaussian distributions. Other tests have been developed to address this limitation \cite{van2005consistency,djuric2010assessment}, but require additional assumptions and approximations which limit their applicability. 

This work presents exact statistical consistency tests for generic Gaussian mixture filters, and by extension Gaussian mixture models, which do not impose any limiting assumptions or additional sampling requirements. Our approach leverages normalized deviation squared (NDS) statistics developed in \cite{ivanov2014evaluating}, which 
(to our best knowledge) have not yet been examined for 
Gaussian mixture pdfs. 
Our main result derives the exact pdf for NDS statistics of an arbitrary finite Gaussian mixture model to be a mixture of generalized chi-square pdfs. This permits formulation of relatively simple statistical hypothesis tests for GM filter pdf consistency checking, using truth model simulations and/or recorded sensor data logs as described above. Although generalized chi-square pdfs cannot be written down in analytical form, they possess closed-form moment-generating functions and permit use of computational tools for accurate evaluation of 
the required quantiles in mixture-based hypothesis testing. We derive the parameters required to obtain the corresponding generalized chi-square mixtures, and use numerical Gaussian mixture estimation examples to demonstrate 
implementation. 
%
In the remainder of the paper: Sec. II describes background and related work; Sec. III derives our main results; Sec. IV describes the numerical example implementations; and Sec. V concludes and discusses avenues for future work. 


%

\section{Background and Related Work} \label{sec:background}
\subsection{Brief Review of GM Filtering}
The GM filter is briefly reviewed in generic probabilistic form, so that the results developed later 
can be directly applied to any finite GM-based estimator for linear or non-linear dynamical systems. 

Let $x_k \in \mathbb{R}^n$ be a random vector representing the unknown discrete time system state at time $k \geq 0$, with dynamics
\begin{align}
    x_{k+1} = f_k(x_{k},u_{k}, w_{k}), \label{eq:dynModel}
\end{align}
for known input $u_{k} \in \mathbb{R}^m$, white process noise input $w_{k} \in \mathbb{R}^{n_w}$, and discrete time evolution function $f_k: \mathbb{R}^n \times \mathbb{R}^m \times \mathbb{R}^{n_w} \mapsto \mathbb{R}^n$. Also, let $y_{k+1} \in \mathbb{R}^{p}$ be a random vector representing observations at time $k+1$, modeled as
\begin{align}
    y_{k+1} = h_{k+1}(x_{k+1},v_{k+1}), \label{eq:measModel}
\end{align}
for white measurement noise process input $v_{k+1} \in \mathbb{R}^{n_v}$ and measurement function $h_{k+1}: \mathbb{R}^n \times \mathbb{R}^{n_v} \mapsto \mathbb{R}^p$. 
Assume that the initial state prior, process, and measurement noise distributions at each time step are given by GM pdfs,
\begin{align}
    p(x_0) &= \sum_{d=1}^{M_{0}} \theta_d \cdot {\cal N}_d(\mu^{d}_0,\Sigma^{d}_0), \label{eq:px0} \\
    p(w_k) &= \sum_{i=1}^{M_{w}} \beta_i \cdot {\cal N}_i(q^{i}_k,Q^{i}_k), \label{eq:pwk} \\
    p(v_{k+1}) &= \sum_{j=1}^{M_{v}} \gamma_j \cdot {\cal N}_j(r^{j}_{k+1},R^{j}_{k+1}), \label{eq:pvk}
\end{align}
where ${\cal N}(a,B)$ denotes the multivariate Gaussian pdf with mean vector $a$ and covariance matrix $B$, and $\theta_d$, $\beta_i$, and $\gamma_j$ represent positive weights which sum to 1 over their respective mixture indices. For linear(ized) models (\ref{eq:dynModel}) and (\ref{eq:measModel}),  
the conditional state transition pdf $p(x_{k+1}|x_{k})$ and observation likelihood pdf $p(y_{k+1}|x_{k+1})$ can also be shown to be GM pdfs \cite{alspach1972nonlinear, schoenberg2012posterior}, 
from which the equations for the discrete time recursive Bayes filter may be developed and simplified. 
Define $Y_{k}=\left\{y_1,\cdots, y_k \right\}$ with $Y_0 \equiv \emptyset$. 
The Bayes filter time update for $k=0,1,2,..$ obtains (via the Chapman-Kolmogorov equation) a GM pdf of the form
\begin{align}
    p(x_{k+1}|Y_k) = \sum_{l=1}^{M_wM_0} \omega_{l,k} \cdot {\cal N}_{l}(\mu^{-,l}_{k+1},\Sigma^{-,l}_{k+1}) . \label{eq:timeUpdateGM}
\end{align}
For the measurement update step, we also have a GM representation for the posterior pdf, via Bayes' rule and the conditional independence of $y_{k+1}$ and $Y_{k}$ given $x_{k+1}$,
\begin{align}
        p(x_{k+1}|Y_{k+1}) &= \frac{ p(x_{k+1}|Y_k) p(y_{k+1}|x_{k+1})}{p(y_{k+1}|Y_{k})} \nonumber \\
        &= \sum_{m=1}^{M_vM_wM_0} \sigma_{m,k+1} \cdot {\cal N}_{l}(\mu^{+,m}_{k+1},\Sigma^{+,m}_{k+1}) .\label{eq:measUpdateGM}
\end{align}
See \cite{alspach1972nonlinear, schoenberg2012posterior} for complete derivations of these expressions. 
The main advantage of  (\ref{eq:timeUpdateGM}) and (\ref{eq:measUpdateGM}) is that they can be found from sets of simpler local Bayesian filters running in parallel, obtaining a natural `divide and conquer' approach to complex filtering problems. 
When (\ref{eq:dynModel}) or (\ref{eq:measModel}) is highly nonlinear, or when $p(x_0)$, $p(w_k)$ or $p(v_{k+1})$ are modeled by more complex non-GM pdfs, GM approximations for (\ref{eq:timeUpdateGM}) and (\ref{eq:measUpdateGM}) can still be obtained by means other than direct function linearization, e.g. using component-wise unscented transforms \cite{faubel2009split}, particle approximations \cite{kotecha2003gaussian}, and other methods \cite{terejanu2011adaptive}. To combat the growth of mixture terms in (\ref{eq:timeUpdateGM}) and (\ref{eq:measUpdateGM}), various condensation methods can be applied to balance computational tractability and information loss \cite{runnalls2007kullback}. 
On the other hand, mixture splitting and hybrid sampling techniques can also be used to dynamically add new terms where needed in the state space to improve approximation accuracy  \cite{stordal2011bridging, havlak2013discrete, raihan2018particle, psiaki2016gaussian}. 
The GM filtering methodology also naturally extends to handling discrete switching dynamics and measurement models, e.g. modeled by jump Markov processes \cite{barshalom2001estimation} and data association uncertainties \cite{wyffels2015negative}. Because of their versatile approximation capabilities, GM filters have attracted interest in a variety of applications, such as orbit determination \cite{psiaki2017gaussian, demars2014collision}, robust navigation \cite{ulmschneider2016gaussian,cui2018robust}, and robotics \cite{ schoenberg2012posterior, burks2021collaborative}. 

The issue examined here is whether the pdfs obtained by a GM filter reasonably reflect the true uncertainties with respect to the state and observations. This issue is central to the filter validation and tuning process, as it determines whether reasonable parameters for dynamics and noise models, mixture condensation/splitting, etc. have been obtained. 

\subsection{Dynamical Consistency and Statistical Validation}
%
An estimator's \emph{dynamic consistency} 
reflects how well the estimator models the true distribution of uncertainty \cite{ivanov2014evaluating}. In the case of Kalman filters for linear(ized)-Gaussian dynamical systems, for instance, dynamic consistency requires that the filter's: (1) estimate of the state be unbiased; (2) estimated state error covariance match the true state error covariance; and (3) innovation errors form a zero mean white noise sequence whose covariance matches the true innovation error covariance \cite{barshalom2001estimation}.
These conditions can be evaluated via chi-square hypothesis tests to examine normalized estimation error squared (NEES) and normalized innovation squared (NIS) statistics. Such statistics serve as `goodness of fit' measures, and the resulting hypothesis tests are useful for tuning filter parameters like process noise covariance \cite{chen2018weak}. 

An open issue for more complex estimation algorithms such as GM filters is that there is no obvious or simple way to validate whether non-Gaussian pdfs (e.g. which may be highly multimodal or skewed)
provide reasonable approximations to state and measurement error uncertainties. 
This is because GM pdfs generally requires higher order moment descriptions beyond their means and covariances. The NEES and NIS statistics for MMSE state estimates derived from a GM-based filter -- i.e. if only mixture mean and mixture covariance are used to form estimates -- do not follow simple chi-square distributions as in the Gaussian case, so classical NEES/NIS chi-square tests are invalid.

Dynamic consistency testing has been extended to general Bayesian filters using variants of other classical goodness of fit tests for arbitrary pdfs. For particle filter estimators, Djuric and M{\'\i}guez considered the Kolmogorov-Smirnov test \cite{djuric2010assessment}, while van der Heijden used discretization-based chi-square tests \cite{van2005consistency}. 
These methods are applicable to quite general and complicated pdfs in theory, though in practice they are also non-trivial to reliably implement. In particular, they require significant parameter tuning and don't scale easily to high dimensions or assessments with multivariate observations. Moreover, they rely on extra sampling steps to compute empirical test statistic distributions, introducing additional sources of approximation error. 

Prior work in statistics and machine learning has examined goodness of fit tests amenable to GM pdfs. 
Refs. \cite{ jitkrittum2017linear,liu2016kernelized} developed techniques based on kernel density estimation, while \cite{wichitchan2022new} developed a  randomized linear projection method for multivariate GMs. 
However, these methods are all tailored to static pdf analysis, and thus are either not applicable or too computationally expensive for dynamic state estimation problems. 
In the signal processing literature, Shah and Li considered Laurent-series based moment-generating function approximations for the pdf of quadratic forms 
of GM-distributed complex random vectors \cite{shah2005distribution}. Their method is not suitable for dynamic consistency testing across multiple time steps, as it is only
applicable only to a single quadratic function evaluation (i.e. at one time step) and does not yield a closed-form cdf for obtaining test thresholds. 

\subsection{Normalized Deviation Squared (NDS) Statistics}
Ivanov, et al. \cite{ivanov2014evaluating} developed several measures of dynamic consistency that make only mild assumptions about the type of information available to an estimator, and are thus theoretically applicable to a broad class of problems involving non-Gaussian pdfs and probabilistic approximation algorithms, including grid and particle methods. In particular, their proposed \emph{normalized deviation squared (NDS)} consistency measure was found to be effective and straightforward to implement via formal hypothesis testing, as described below. 

The NDS consistency criterion states that an $n-$variate random variable $\hat{x}$ with mean $\bar{\mu} \in \mathbb{R}^n$ and covariance matrix $\bar{\Sigma} \in \mathbb{R}^{n \times n}$ is a consistent estimate of the random variable $x \in \mathbb{R}^n$ if
\begin{align}
    \forall \epsilon>0: \  &P\left\{(x-\bar{\mu})^T \bar{\Sigma}^{-1}(x-\bar{\mu}) \leq \epsilon \right\} \nonumber \\ 
    &\geq P\left\{(\hat{x}-\bar{\mu})^T \bar{\Sigma}^{-1}(\hat{x}-\bar{\mu}) \leq \epsilon \right\}.
\end{align}
Intuitively, this condition `implies that the concentration ellipse of any probability mass of $\hat{x}$ contains a larger or equal probability mass of $x$' \cite{ivanov2014evaluating}. Ivanov, et al. develop the following hypothesis test for consistency with respect to an independent set of estimates $\hat{X}_M = \left\{ \hat{x}_1,...,\hat{x}_M \right\}$ and random vectors of interest $X = \left\{x_1,...,x_M\right\}$, for $M \geq 1$. 
Let $H_0$ be the null hypothesis corresponding to the case where $\hat{X}_M$ are all NDS consistent, and define the test statistics for $c=1,...,M$
\begin{align}
    &Q(X) = \sum_{c=1}^{M}q_c(x_c), \ \ \  \hat{Q}(\hat{X}) = \sum_{c=1}^{M}\hat{q}_c(\hat{x}_c), \label{eq:NDSstatsdef} \\
    &q_c(x_c) = (x_c-\bar{\mu}_c)^T\bar{\Sigma}^{-1}_c(x_c-\bar{\mu}_c), \label{eq: obsndsstat} \\
    &\hat{q}_c(\hat{x}_c) = (\hat{x}_c-\bar{\mu}_c)^T\bar{\Sigma}^{-1}_c(\hat{x}_c-\bar{\mu}_c). \label{eq:predndsstat}
\end{align}
Then, if $H_0$ is true, it can be shown from the definition of the NDS criterion that \cite{ivanov_MS_thesis} 
\begin{align}
    &\forall \epsilon_c >0, \ P\left\{q_c(x_c) \leq \epsilon_c \right\} \geq P\left\{\hat{q}_c(\hat{x}_c) \leq \epsilon_c \right\} \\
    \Rightarrow &\forall \epsilon > 0, \ P\left\{Q(X) \leq \epsilon \right\} \geq P\left\{\hat{Q}(\hat{X}) \leq \epsilon \right\}.
\end{align}
For a desired significance level $\alpha = P(\mbox{reject } H_0|H_0 \mbox{ true})$, the critical region $\rho = [\tau,+\infty)$ for the observed statistic $Q(X)$ can be found from the cumulative distribution function (cdf) of $\hat{Q}(\hat{X})$, where
\begin{align}
    P(Q(X) \in \rho) \leq P(\hat{Q}(\hat{X}) \in \rho) \leq \alpha. \label{eq: nds_test_alpha}
\end{align}
Hence, the null hypothesis $H_0$ for NDS consistency of $\hat{X}$ can be rejected at significance level $\alpha$ if the observed $Q(x) \in \rho$. Note that Ivanov, et al. suggest using $M > 1$ to maximize the test' s statistical power, i.e. $P(\mbox{reject } H_0|H_0 \mbox{ false})$. Although ensuring complete independence of the estimates in $\hat{X}$ is not always possible if they are taken from the same dynamic process sequence, accurate and reliable results are obtainable as long as $\hat{x}_1,...,\hat{x}_M$ are spaced far enough apart in time, e.g. as to achieve very low correlation. Moreover, $\hat{X}$ and $X$ may be respectively replaced with predicted and actual observation sequences $\hat{Y}$ and $Y$ to derive a separate NDS hypothesis test with respect to recorded sensor data logs. 


For GM filter evaluation, the NDS consistency test leverages the mixture mean vector and covariance matrix at selected time steps $c=1,...,M$, which are easily computed for GM pdfs $p(\hat{x}_c)$. Also, the test advantageously remains valid even if mixture splitting/compression methods are applied to $p(\hat{x}_c)$ at any instance $c$. However, the cdf $P(\hat{Q}(\hat{X}))$ is generally non-trivial to obtain if each $p(\hat{x}_c)$ is an arbitrary $n$-dimensional GM pdf. 
Ivanov, et al. do not provide exact results for $P(\hat{Q}(\hat{X}))$ in any particular case, other than pointing out that the NDS hypothesis test reduces to the classical chi-square NEES test in the special case that $p(\hat{x}_c)$ is Gaussian. 
The most direct way to address this issue is to sample the GM pdfs $p(\hat{x}_c)$ generated by the filter and estimate the empirical cdf $\hat{P}(\hat{Q}(\hat{X}))$ via (\ref{eq:NDSstatsdef}) and (\ref{eq:predndsstat}). 
The empirical cdf can then be used to approximate the critical region $\rho$ for the desired $\alpha$. 
Although conceptually simple and straightforward, such a sampling process can become expensive for large $n$ and $M$, and is unreliable if the sample size is too small. For example, even if $n=1$, the presence of many low-weight but non-negligible tail masses in the GMs may require hundreds of thousands or millions of samples to estimate $\rho$ with reasonable accuracy. 
To address this issue, we next derive exact expressions for the pdfs $p(\hat{q}_c(\hat{x}_c))$ and $p(\hat{Q}(\hat{X}))$, which in turn allow us to obtain and interrogate their exact cdfs $P(\hat{q}_c(\hat{x}_c))$ and $P(\hat{Q}(\hat{X}))$ for more accurate and stable specification of $\rho$. 

\section{Distributions of NDS Statistics for GMs} \label{sec:method}
%
\subsection{Gaussian NDS Statistics}
Before deriving the main results, it will be useful to first examine the exact pdf of the NDS statistic for an arbitrary Gaussian random vector. Ignoring time indices for convenience, suppose $x \in \mathbb{R}^n$ has pdf $p(x) = {\cal N}(\mu,\Sigma)$ with mean $\mu \in \mathbb{R}^n$ and positive definite symmetric covariance $\Sigma \in \mathbb{R}^{n \times n}$. 
Then the NDS statistic $q(x) \in \mathbb{R}$ is
\begin{align}
    q(x) &= (x-\mu)^T \Sigma^{-1}(x-\mu) \\
    &=x^T\Sigma^{-1}x -\mu^T\Sigma^{-1}x - x^T\Sigma^{-1}\mu + \mu^T\Sigma^{-1}\mu \nonumber \\
    &= x^T\Sigma^{-1}x - 2\mu^T\Sigma^{-1}x + \mu^T\Sigma^{-1}\mu \nonumber \\
    \rightarrow q(x) &=x^TAx + q_1^T x + q_0, \label{eq: quadformgen}
\end{align}
where $A=\Sigma^{-1}$, $q_1 = -2 A \mu$, and $q_0 = \mu^T A \mu$. From refs. \cite{mathai1992quadratic} (p. 28) and \cite{das2021method}, it follows that $q(x)$ can be expressed as a sum of non-central chi-square ($\chi^2$) random variables and a Gaussian scalar random variable. 
We apply a slightly modified version of the derivation from \cite{das2021method} to show this. Let $S$ be the matrix square root of $\Sigma$, such that $\Sigma = SS^T$, and define $z=S^{-1}(x-\mu)$, so that 
\begin{align}
    \tilde{q}(z) = z^T \tilde{A} z + \tilde{q}_1^T z + \tilde{q}_0,
\end{align}
with $\tilde{A} = S^T A S$, $\tilde{q}_1 = 2S^TA\mu + S^Tq_1$, and $\tilde{q}_0 = q(\mu) = \mu^TA\mu + q_1^T\mu + q_0$. 

Next, take the eigen-decomposition $\tilde{A} = PDP^T$, where $PP^T = P^TP = I$ and $D = \mbox{diag}[D_1,...,D_n]$ is the diagonal matrix of eigenvalues of $\tilde{A}$. Let $s=P^Tz$, whence it can be easily shown that $p(s)={\cal N}(0,I)$ and we obtain
\begin{align}
    \hat{q}(s) &= s^TDs + b^Ts + \tilde{q}_0, \\
    \mbox{where \ } b &= P^T \tilde{q}_1 = 2P^T S^T A \mu + P^T S^T q_1. \label{eq: qhat_s1}
\end{align}
Since $A=\Sigma^{-1}$ has no zero eigenvalues, $\tilde{A}$ also has no zero eigenvalues, so that $\hat{q}(s)$ can be expressed in scalar form as
\begin{align}
    &\hat{q}(s) = \sum_{i=1}^{n} \left(D_is_i^2 + b_is_i \right) + \tilde{q}_0,  \nonumber \\
    &= \sum_{i=1}^{n} D_i \left(s_i + \frac{b_i}{2D_i} \right)^2 - \sum_{i=1}^{n}D_i\left(\frac{b_i}{2D_i} \right)^2 + \tilde{q}_0, \label{eq:ssum}
\end{align}
where $i$ indexes over the eigenvalues in $D$ and corresponding elements in vectors $s$ and $b$. Since the first term in (\ref{eq:ssum}) is a sum of squares of independent Gaussian variables with unity variance and non-zero means, (\ref{eq:ssum}) is equivalent to
\begin{align}
\hat{q}(s) = \sum_{i=1}^{n} D_i r_i + t, \label{eq: qhat_s2}
\end{align}
where $r_i \in \mathbb{R}$ is a non-central $\chi^2$ random variable with 1 degree of freedom and non-centrality parameter $(b_i/2D_i)^2$, denoted $r_i \sim \chi^{2,n.c.}_{(b_i/2D_i)^2}$. 
Also, $t=q(\mu)-w^T\lambda$, where $w$ is the vector of the \emph{unique} eigenvalue entries in $D$, and $\lambda$ is a vector of non-centrality parameters with the same length as $w$, with corresponding elements $j$ determined as
\begin{align}
    \lambda_j = \frac{1}{4w_j^2} \sum_{i:D_i=w_j}b_i^2, 
\end{align}
where the sum is over elements with the same eigenvalue (algebraic multiplicity greater than 1).
Define $k$ as the degree of freedom vector, also of the same length as $w$ and $\lambda$, where each element of $k$ gives the algebraic multiplicity of the unique eigenvalue entries in $D$. 
From all this, it finally follows that $\hat{q}(s)$ has a \emph{generalized chi-square distribution}, denoted by the four parameter notation
\begin{align}
    \hat{q}(s) &\sim \tilde{\chi}^2_{w,k,\lambda,t} = p(q(x)).
    \label{eq: genchisquaredef}
\end{align}
Note in the special case where $\mu=0$ and $\Sigma=I$, this result reduces to a `simple' $\chi^2$ distribution with $n$ degrees of freedom, which corresponds to the pdf for NEES and NIS statistics in Kalman filter consistency tests. Furthermore, with only minor changes, eq. (\ref{eq: genchisquaredef}) is the pdf for a general quadratic form of Gaussian random variable given by $q'(x) = x^TAx + q'^T_1x + q'_0$, where $A$ is \emph{any} symmetric matrix, and $q'_1$ and $q'_0$ are arbitrary; in this case, the zero and non-zero eigenvalues of $\tilde{A}$ in the derivation above will contribute and $t$ changes to a scalar Gaussian random variable $t'$ with mean $t$ and non-zero standard deviation (see \cite{das2021method} for details). 

\subsection{Single Gaussian Mixture NDS Statistics}
Now suppose $x$ has a GM pdf $p(x)$ with mixture mean $\bar{\mu}$ and positive definite mixture covariance matrix $\bar{\Sigma}$,
\begin{align}
    &p(x)=\sum_{g=1}^{G}\eta_g \cdot {\cal N}(\mu_g,\Sigma_g), \\
    &\bar{\mu} = \sum_{g=1}^{G}\eta_g \mu_g, \label{eq: mixmean}, \ \
    \bar{\Sigma} = \sum_{g=1}^{G}\eta_g (\Sigma_g + \mu_g\mu_g^T) - \bar{\mu}\bar{\mu}^T 
\end{align}
We want the pdf for the NDS statistic
\begin{align}
    q(x) = (x-\bar{\mu})^T \bar{\Sigma}^{-1}(x-\bar{\mu}). \label{eq:nds_gm_def}
\end{align}
To obtain $p(q(x))$ in this case, start with the joint distribution $p(x,l)$, where $l \in \left\{1,2,...,G \right\}$ is a discrete random variable which indexes a particular mixture component. The marginal $p(x) = \sum_{g=1} p(x,l=g) = \sum_{g} p(x|l=g)P(l=g)$ is obtained as a GM with $p(x|l=g) = {\cal N}(\mu_g,\Sigma_g)$ and $P(l=g) = \eta_g$. 
Then define the $l$-indexed NDS statistic 
\begin{align}
    q(x,l=g) = (x_g-\bar{\mu})^T \bar{\Sigma}^{-1}(x_g-\bar{\mu}),
\end{align}
where $x_g \sim p(x|l=g) = {\cal N}(\mu_g,\Sigma_g)$. Letting $A=\bar{\Sigma}^{-1}$, $q_1 = -2A\bar{\mu}$, and $q_0 = \bar{\mu}^TA\bar{\mu}$, an expression similar to eq. (\ref{eq: quadformgen}) is obtained,
\begin{align}
    q(x,l=g) = x_g^TAx_g + q_1^Tx_g+ q_0. \label{eq: NDS_quadform}
\end{align}
Now from the chain rule of probability, we have
\begin{align}
p(q(x,l=g),g) = p(q(x,l=g)|l=g)P(l=g). \label{eq:chain_rule}
\end{align}
The previous logic for the single Gaussian case is now carefully adapted as follows to determine $p(q(x,l=g)|l=g)$ for each $g=1,...,G$. First, take $S_g$ to be the square root matrix of $\Sigma_g$ such that $\Sigma_g = S_gS_g^T$, and define $z_g = S_g^{-1}(x-\mu_g)$, so that 
\begin{align}
    \tilde{q}(z,l=g) = z_g^T\tilde{A}z_g + \tilde{q}^T_{1,g} z_g + \tilde{q}_{0,g},
\end{align}
where $\tilde{A}_g = S^T_g A S_g$, $\tilde{q}_{1,g} = 2S^T_gA\mu_g + S^T_g q_1$, and $\tilde{q}_{0,g} = \mu^T_g A \mu_g + q_1^T \mu_g + q_0$. Then, proceeding with the eigen-decomposition of $\tilde{A}_g = P_gD_gP_g^T$ and defining $s_g = P^T_g z_g$, analogous expressions for eqs. (\ref{eq: qhat_s1})-(\ref{eq: qhat_s2}) are obtained with respect to summation over the $n_g$ eigenvalues of $\tilde{A}_g$. In particular, we obtain
\begin{align}
    \hat{q}(s,l=g) &= s_g^TDs_g + b^T_g s_g + \tilde{q}_{0,g}, \\
    \mbox{where \ } b_g &= P^T \tilde{q}_{1,g} = 2P_g^T S_g^T A \mu_g + P_g^T S_g^T q_{1}. \label{eq: qhat_s1_g} \\
\rightarrow \hat{q}(s,l=g) &= \sum_{i=1}^{n_g}D_{i,g}r_{i,g} + t_g, \label{eq: qhat_s2_g}   
\end{align}
where $r_{i,g} \sim \chi^{2,n.c.}_{(b_{i,g}/2D_{i,g})^2}$ and $t_g = q(\mu_g) - w^T_g \lambda_g$ with $w_g$ and $\lambda_g$ defined similarly as before, except now in reference to the unique eigenvalue entries $D_{i,g}$ of the diagonal matrix $D_g$. Letting $k_g$ be the vector of algebraic multiplicities for the entries of $D_g$, it thus follows that $p(q(x,l=g)|l=g)$ is a generalized chi-square pdf with parameters $(w_g,k_g,\lambda_g,t_g)$, 
\begin{align}
    \hat{q}(s,l=g) \sim \tilde{\chi}^2_{w_g,k_g,\lambda_g,t_g} = p(q(x,l=g)|l=g).
\end{align}
Note that a subtle but crucial difference from the earlier NDS pdf derivation in the single Gaussian case is that the \emph{mixture} statistics $\bar{\mu}$ and $\bar{\Sigma}$ are used to define the $l$-indexed NDS statistic, \emph{not} the individual component statistics $\mu_g$ and $\Sigma_g$ (although the latter still appear in certain places). Plugging this result and $P(l=g) = \eta_g$ into (\ref{eq:chain_rule}), 
\begin{align}
    p(q(x,l=g),l=g) = \eta_g \cdot \tilde{\chi}^2_{w_g,k_g,\lambda_g,t_g}.
\end{align}
Finally, marginalizing the component index $l$ gives a mixture of generalized chi-square pdfs,
\begin{align}
    p(q(x)) &= \sum_{g=1}^{G} p(q(x,l=g)|l=g)P(l=g) \\
    &= \sum_{g=1}^{G} \eta_g \cdot \tilde{\chi}^2_{w_g,k_g,\lambda_g,t_g}. \label{eq:mix_genx2}
\end{align}

\subsection{Sums of Independent GM NDS Statistics}
Let $x_1,...,x_M$ be $M$ independent random vectors with GM pdfs $p(x_1),...,p(x_M)$, which for $c \in \left\{ 1,...,M \right\}$ are 
\begin{align}
    p(x_c) = \sum_{g_c=1}^{G_c}w_{g_c}{\cal N}(\mu_{g_c},\Sigma_{g_c}),
\end{align}
with corresponding mixture means and covariances $\bar{\mu}_c$ and $\bar{\Sigma}_c$, respectively, defined as in (\ref{eq: mixmean}). Defining $q_c(x_c)$ via (\ref{eq:nds_gm_def}) with $x=x_c$, $\bar{\mu}=\bar{\mu}_c$ and $\bar{\Sigma}=\bar{\Sigma}_c$, we want to find the pdf for the sum of NDS statistics
\begin{align}
    Q(x_{1:M}) = \sum_{c=1}^{M}q_c(x_c).
\end{align}
Recognizing from independence that $p(x_1,...,x_M) = \prod_{c=1}^{M}p(x_c)$, the joint pdf over $x_1,...,x_M$ is a product of all $M$ GMs, which can be expressed as a `mixture of mixtures',
\begin{align}
    p(x_1,...,x_M) = \sum_{g_1=1}^{G_1} \cdots \sum_{g_M=1}^{G_M} \prod_{c=1}^{M}w_{g_c} {\cal N}(x_c;\mu_{g_c},\Sigma_{g_c}). \nonumber 
\end{align}
Following similar logic as for the single GM case, this mixture can be induced via marginalization over a single discrete `super index' variable $l^{*}$, whose each realization selects exactly one mixture component index from each GM $p(x_c)$. If $l^*$ is an $M$-dimensional vector where $l^*(c)=l_c \in \left\{1,...,G_c\right\}$, then $l^*$ has $G^{\#} = \prod_{c=1}^{M}G_c$ possible realizations and the joint distribution $p(x_1,...,x_M,l^*)$ is
\begin{align}
    p(x_1,&...,x_M,l^*=[l_1,...,l_M]) = w_{l*} \cdot {\cal N}(\vec{x}; \vec{\mu}_{l^*},\mathbf{\Sigma}_{l^*}), \nonumber \\ 
    &\mbox{where \ } w_{l*} = \prod_{c=1}^{M}w_{l_c} = P(l^*=[l_1,...,l_M]) \\  
    &\vec{\mu}_{l^*} = \begin{bmatrix} 
    \mu_{l_1}, & \cdots &, \mu_{l_M}
    \end{bmatrix}^T, \\ 
    &\mathbf{\Sigma}_{l^*} = \mbox{blkdiag}[\Sigma_{l_1},\cdots,\Sigma_{l_M}].
\end{align}
Thus, finding $p(Q(x_{1:M}))=p(Q(\vec{x}))$ is the same as finding $p(q(x))$ for the single GM case above, where $x=\vec{x}$ is an $nM$-dimensional vector whose GM pdf has $G^{\#}$ components. As such, $p(Q(\vec{x}))$ is exactly a mixture of generalized chi-square pdfs with $G^{\#}$ components. Note that the  constants for the corresponding quadratic form expansion in (\ref{eq: NDS_quadform}) are 
\begin{align}
    A = \mathbf{A} &= \mbox{blkdiag}[A_1,...,A_M], \ \ A_c = \bar{\Sigma}_c^{-1}  \\
    q_1 &= \vec{q}_1 = -2\mathbf{A}\bar{\mu}, \ \
    q_0 = \bar{\mu}^T \mathbf{A} \bar{\mu}\\
    \bar{\mu}^T &= \begin{bmatrix}
        \bar{\mu}_1^T, & \cdots, &\bar{\mu}_M^T 
    \end{bmatrix}.
\end{align}

\subsection{Computational Implementation and Considerations}
Generalized chi-square pdfs and cdfs do not possess simple closed-form expressions, although they do possess closed-form moment-generating functions. Fortunately, computational tools are available to evaluate the generalized chi-square pdfs and cdfs, as well as sample from them. Our implementations use the Generalized Chi-square Toolbox for Matlab \cite{das2021method} to evaluate the cdf of (\ref{eq:mix_genx2}) as a weighted sum of the component term cdfs, which are found via the {\ttfamily gx2cdf} command\footnote{we found using the {{\ttfamily gx2cdf\_ruben}} command with 200 evaluation points improves accuracy for $\alpha \leq 0.1$ when elements of the non-centrality parameter $\lambda$ are relatively large}. To evaluate critical region $\rho=[\tau,+\infty)$ for a given confidence level $\alpha$ per (\ref{eq: nds_test_alpha}), we use Matlab's {\ttfamily fminsearch} command to determine $\tau$ minimizing $|\int_{0}^{\tau}p(q(x))dq(x)-\alpha|$ to within a specified tolerance.  

For practical GM filtering problems, the potentially explosive growth of $G=G^{\#}$ in (\ref{eq:mix_genx2}) is another another major issue to contend with when evaluating sums of NDS statistics for $M>1$.  
Even if the number of posterior Gaussian terms $G_1,...,G_M$ remains bounded through repeated mixture condensation to relatively small values, the number of cdf terms to evaluate can still become quite large, e.g. for $G_1=G_2=...G_M=5$, with $M=10$ independent steps one obtains $G^{\#}=9,765,625$. To deal with this, we have found weight-based retention of only the top $G^{*} << G^{\#}$ terms in (\ref{eq:mix_genx2}) effective at maintaining tractability. 
Future work will consider a wider possible range of techniques. 


\section{Simulation Results} \label{sec:results}
\subsection{Static GM validation}
\begin{figure*}
    \subfigure[]{
    \includegraphics[width=6.5cm]{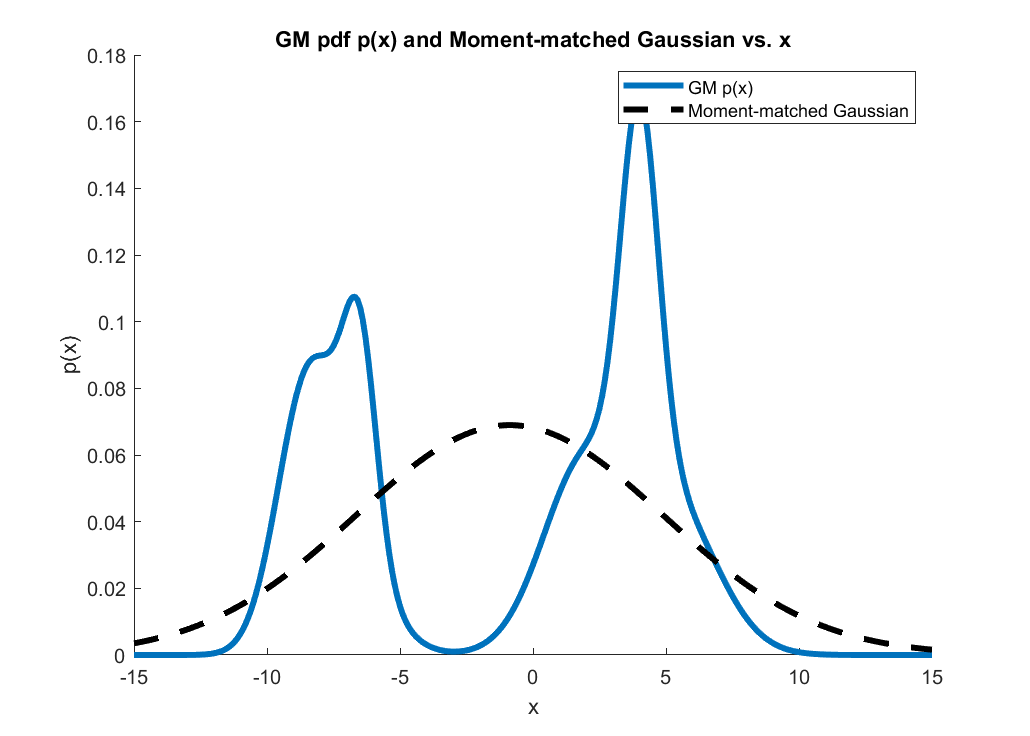}
    }
    \hspace{-1cm}
    \subfigure[]{
    \includegraphics[width=6.5cm]{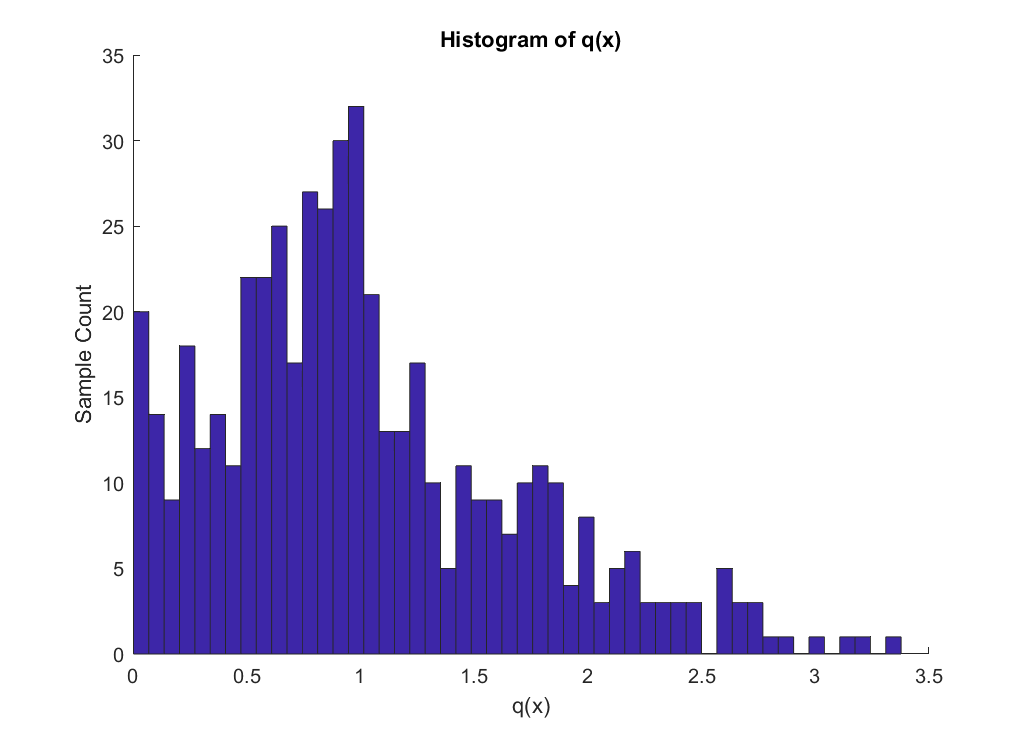}
    }
    \hspace{-1cm}
    \subfigure[]{
    \includegraphics[width=6.5cm]{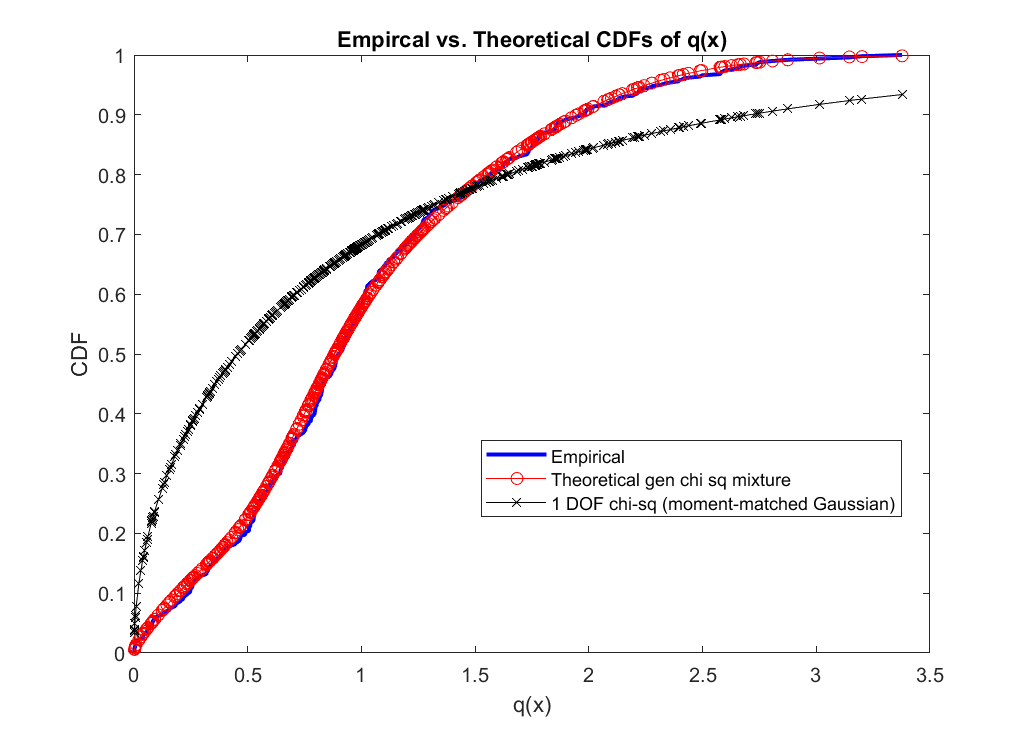}
    }
    \caption{(a) multi-modal GM pdf $p(x)$ with moment-matched Gaussian; (b) $q(x)$ histogram for 500 samples; (c) empirical vs. theoretical $q(x)$ CDFs, showing strong agreement between empirical and generalized chi-squared mixture, and strong disagreement with 1 DOF simple chi-square ($q(x)$ pdf for moment-matched Gaussian).}
    \label{fig:scalarGM}
\end{figure*}

Fig. \ref{fig:scalarGM}(a) shows a scalar GM pdf $p(x)$ and a moment-matched Gaussian pdf (black), which might be naively used to generate a simpler approximate CDF for the NDS statistic $q(x)$ to perform consistency testing in place of its true CDF. 500 Monte Carlo samples are used to evaluate $q(x)$, whose histogram in Fig. \ref{fig:scalarGM}(b) displays strongly skewed and multimodal features. In Fig. \ref{fig:scalarGM}(c), there is nearly perfect agreement between the empirical CDF for $q(x)$ and the theoretically predicted generalized chi-square mixture pdf (\ref{eq:mix_genx2}), but strong divergence from the 1 degree of freedom chi-square CDF obtained by evaluating $q(x)$ via the naive moment-matched Gaussian approximation. 

\begin{figure}
    \centering
    \includegraphics[width=7.5cm]{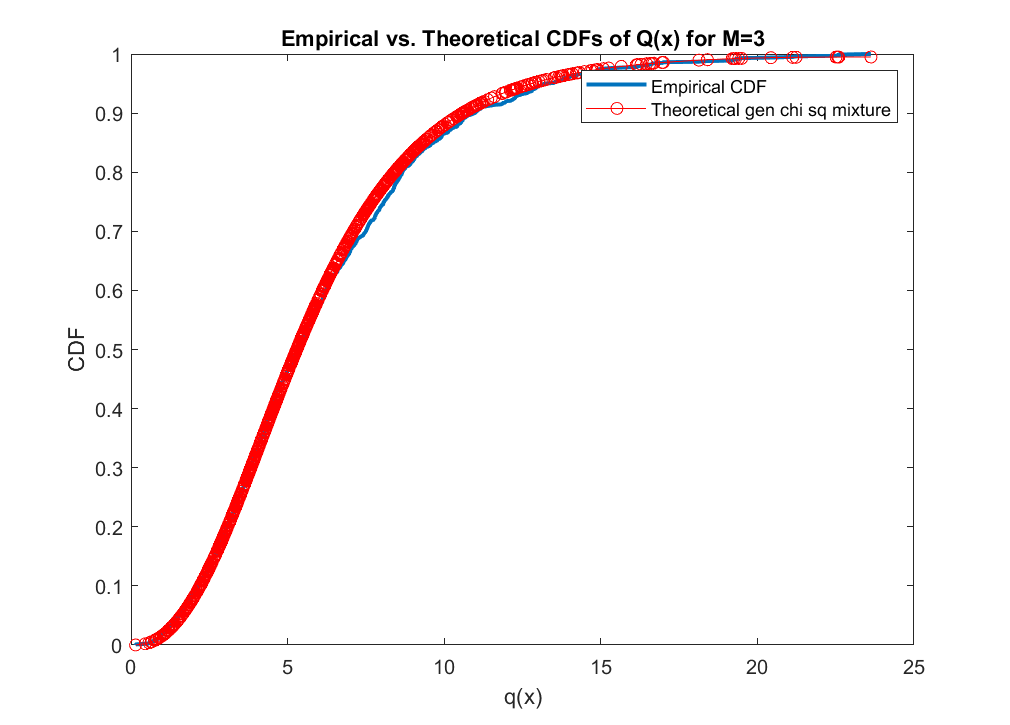}
    \caption{Empirical CDF from 1000 samples of $Q(x)$ for $M=3$ multivariate GMs in $n=2$ dimensions. 
    }
    \label{fig:vectorGM}
    \vspace{-0.75 cm}
\end{figure}

Fig. \ref{fig:vectorGM} also shows good agreement between the empirical and theoretical CDFs for a sum of NDS statistics $Q(x_{1:M})$ for $M=3$ multivariate GM pdfs. In this example, there are $G_1=G_2=G_3=3$ terms for each independent GM $p(x_1)$, $p(x_2)$, and $p(x_3)$ in $n=2$ dimensions (generated with random means, weights, and covariance matrices). 
In this case, the critical region threshold for $\alpha=0.05$ is calculated by applying {\ttfamily fminsearch} to the cdf of the $G^{\#}=27$ term mixture pdf (\ref{eq:mix_genx2}) with a tolerance of $1$E$-06$, resulting in $\tau=12.7809$. Approximately $5\%$ of the 1000 samples from the GMs $p(x_{1:3})$ produce $Q(x_{1:3})>\tau$, validating the result. 

\subsection{Dynamic GM filter validation}
\begin{figure*}
    \centering
    \subfigure[]{
    \includegraphics[width=5cm]{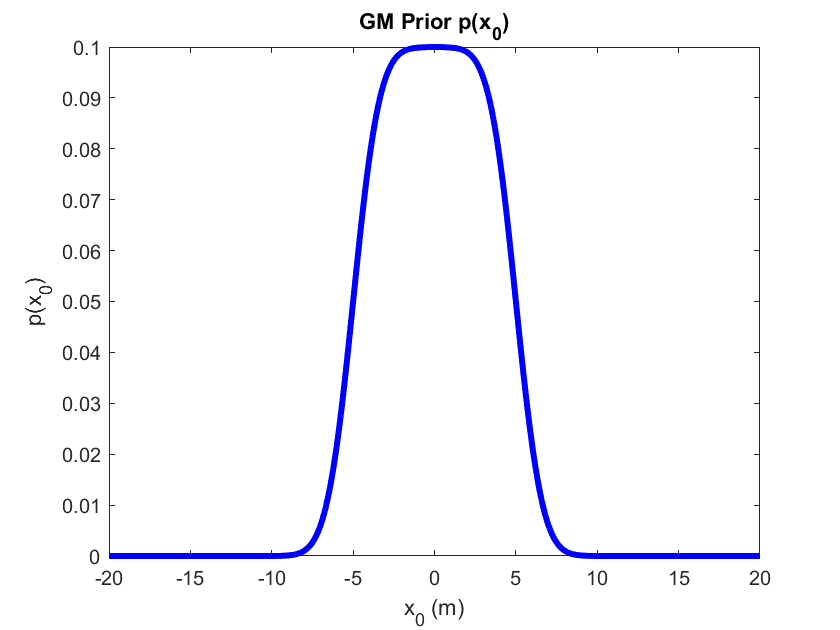}
    }
    \hspace{-1.3cm}
    \subfigure[]{
    \includegraphics[width=5cm]{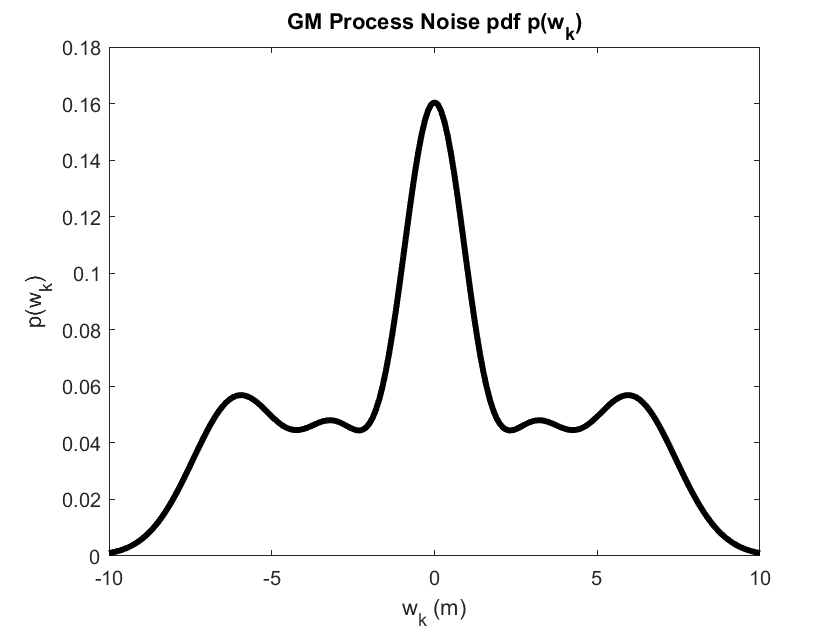}
    }
    \hspace{-1.3cm}
    \subfigure[]{
    \includegraphics[width=5cm]{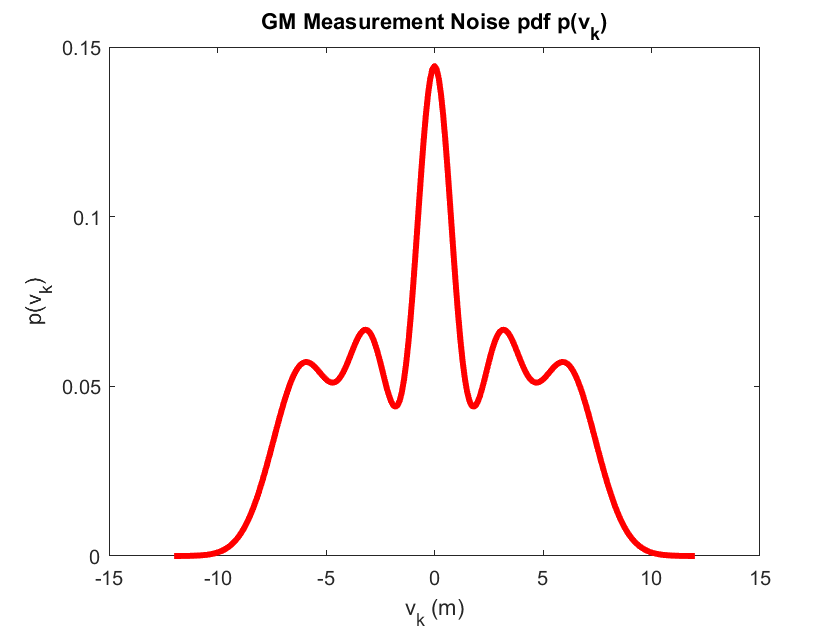}
    }
    \hspace{-1.3cm}
    \subfigure[]{
    \includegraphics[width=5cm]{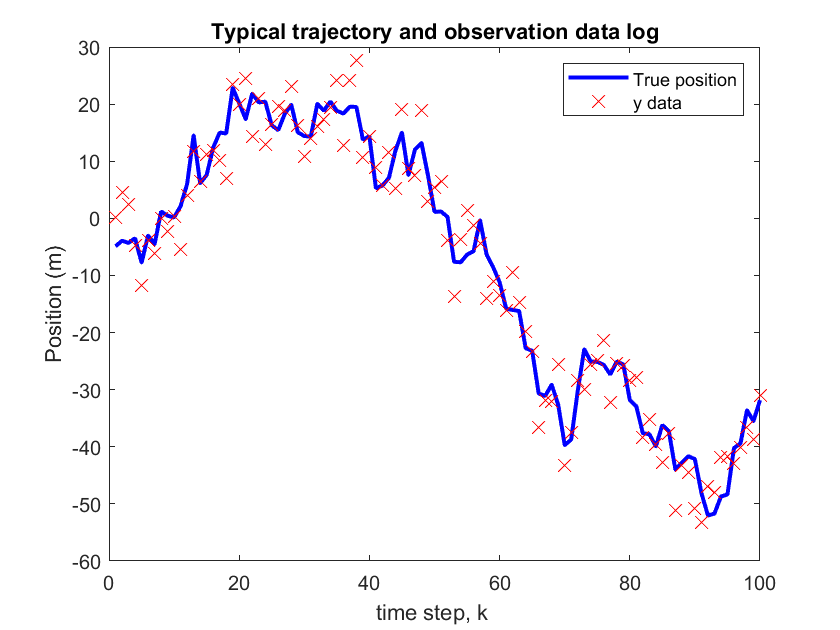}
    } \\
    \subfigure[]{
    \includegraphics[width=5cm]{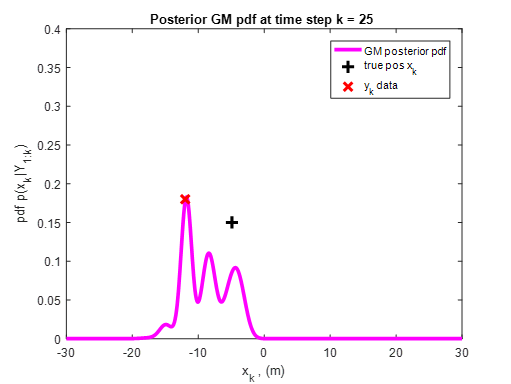}
    }
    \hspace{-1.3cm}
    \subfigure[]{
    \includegraphics[width=5cm]{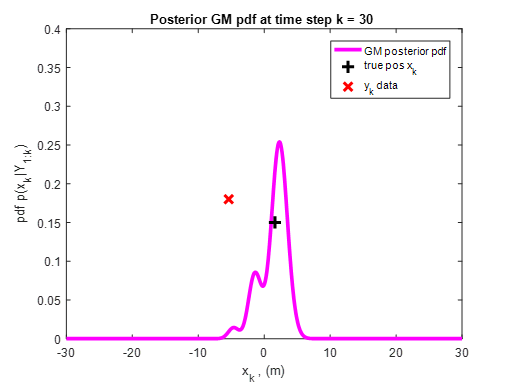}
    }
    \hspace{-1.3cm}
    \subfigure[]{
    \includegraphics[width=5cm]{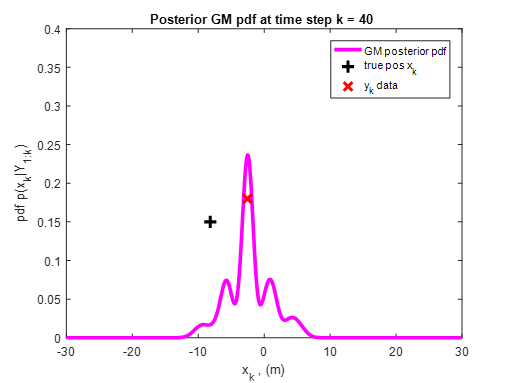}
    }
    \hspace{-1.3cm}
    \subfigure[]{
    \includegraphics[width=5cm]{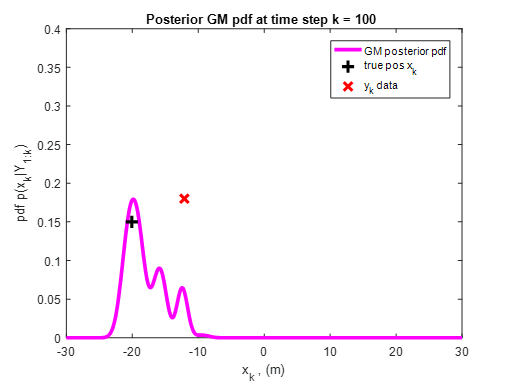}
    }
    \caption{GM priors and typical data for 1D localization example: (a) prior $p(x_0)$; (b) process noise pdf $p(w_k)$; (c) measurement noise pdf $p(v_k)$; (d) ground truth sample trajectory and measurements. (e)-(f) GM filter-estimated posterior pdf (magenta lines) at selected times ($k=25,30,40,100$), showing true platform state (black cross) and $y_k$ data (red x). } 
    \label{fig:gmFilterSetup}
\end{figure*}

We demonstrate GM filter dynamic consistency testing for a 1D localization problem with non-Gaussian noise. 
The platform position $x_k$ at discrete time $k=0,1,2,...$ is modeled with a GM prior $p(x_0)$ and follows a simple random walk process with i.i.d. process noise $w_k$, which also follows a GM pdf $p(w_k)$. The measurements are given by a GPS-like transponder with i.i.d. measurement error $v_k$, which is affected by multi-modal multi-path signal reflections in an urban canyon modeled by GM $p(v_{k+1})$.
The linear random walk dynamics and measurement models are 
\begin{align}
    x_{k+1} = x_k + w_k, \ \ y_{k+1}= x_{k+1} + v_{k+1},
\end{align}
where (following (\ref{eq:px0})-(\ref{eq:pvk})) 
the GM pdfs with for 
$p(x_0)$ ($M_0=5$), 
$p(w_k)$ ($M_w=5$), and 
$p(v_{k+1})$ ($M_v=5$) 
are 
shown in Fig. \ref{fig:gmFilterSetup}(a)-(c). 
Sample ground truth platform position and measurement data generated from this model are shown in Fig. \ref{fig:gmFilterSetup}(d). 
Fig. \ref{fig:gmFilterSetup}(e)-(h) show the resulting posterior GM pdfs produced by the filter using this data log with $G_k=10$ components per time step, following reduction from $G_{k-1}=M_wM_v$ components via mixture condensation \cite{runnalls2007kullback} with $G_{0}=M_0$. The GM filter updates the mixture weights $\sigma_{m,k}$ in the posterior (\ref{eq:measUpdateGM}) in response to received $y_k$, and generally manages to track the true platform location. However, $y_k$ often randomly `spikes' due to $p(v_{k+1})$'s multi-modality, making this a tricky problem even when the filter uses the true process parameters (e.g. note how the true $x_{k}$ lies in smaller modes in Fig. \ref{fig:gmFilterSetup}(e) and (g)).  

\begin{figure}
    \centering
    \hspace{-0.4cm}
    \subfigure[]{
    \includegraphics[width=4.35cm]{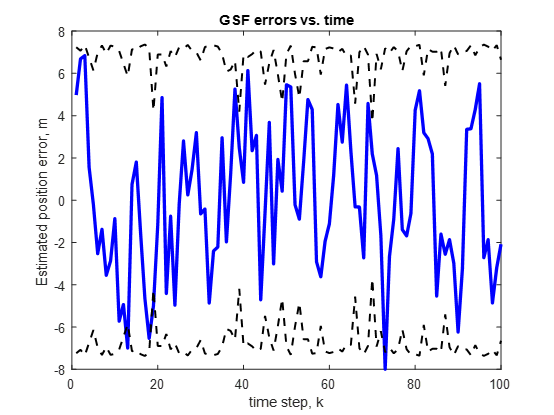}
    }
    \hspace{-0.8cm}
    \subfigure[]{
    \includegraphics[width=4.35cm]{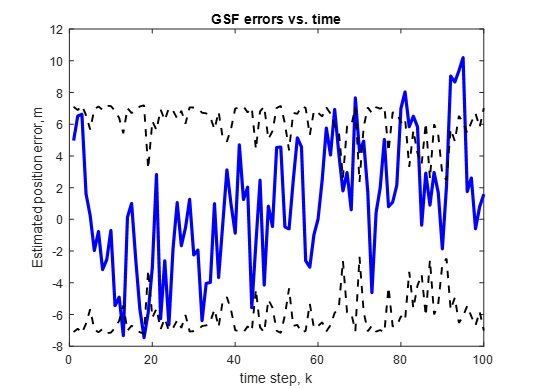}
    }
    \caption{GM filter MMSE estimation error (blue) and $2\sigma$ (black dash) for (a) nominal and (b) mismatched cases. }
    \label{fig:gmFilter_SquaredErrors}
    \vspace{-0.5 cm}
\end{figure}

We consider NDS testing with respect to $x_k$ for the nominal filter model and for a mismatched  model. The MMSE state estimation error plots and corresponding $2\sigma$ bounds for each model case (obtained from the GM pdf mixture mean $\bar{\mu}_k$ and covariance $\bar{\Sigma}_k$ at each time step) are respectively shown in Figure \ref{fig:gmFilter_SquaredErrors} (a) and (b).  

In the nominal case, we assessed $Q(x_{1:M})$ for $M=15$ sufficiently separated time steps, $k=$5:15:75, where analysis of the state estimation error time series showed autocorrelations $<0.02$ for $\Delta k \geq 5$ time steps apart. With these values, we obtained a critical region threshold for $\alpha=0.01$ at $\tau= 28.8442$, while the observed NDS statistic was $Q(x_{1:M}) = 18.5622$. Hence, $H_0$ is not rejected for $p(x_k)$, and the GM filter's pdfs $p(x_k|Y_k)$ appear NDS consistent. 

In the mismatched model case, the filter dynamics incorrectly models the random walk motion as $x_{k+1}= 0.5x_k + w_k$. Analysis of the state error shows that $\Delta k \geq 30$ yields autocorrelations $<0.02$, so in this case we assessed $Q(x_{1:M})$ for $M=4$ separated time steps at $k=5, 35, 65, 95$. For $\alpha=0.01$, $\tau$= 12.2645; with the observed $Q(x) = 15.5434$, $H_0$ can be rejected and we can declare $p(x_k|Y_k)$ to not be NDS consistent at the 1\% significance level. This confirms the results observed in Fig. \ref{fig:gmFilter_SquaredErrors} (b), where errors remain close to the bounds at the start but then gradually wander outside the opposite bounds toward the end. 

\section{Conclusion and Future Work} \label{sec:concl}
This work derived the exact distributions for normalized deviation squared (NDS) statistics for arbitrary multivariate Gaussian mixture (GM) pdfs in the form of mixtures of generalized chi-square pdfs. When used in conjunction with readily available computational tools, the results reliably and accurately determine the critical regions for NDS-based dynamic consistency hypothesis tests for GM filters. This provides a useful tool for formal statistical filter validation and tuning in many different state estimation applications involving non-Gaussian uncertainties. 
Numerical examples validated the results on static GMs, and demonstrated utility in dynamic state estimation error consistency checking for a GM filter. Though not shown here, the results also extend to validation of GM filter measurement model pdfs $p(y_k|Y_{k-1})$ (which are also produced as GMs).

Next steps for future work include formal comparison of the NDS test to other statistical tests developed for non-Gaussian filtering \cite{van2005consistency,djuric2010assessment}, as well as examination of other computationally efficient and accurate strategies for coping with mixture size explosion for testing large multi-time step sample sizes. Since the ellipsoidal probability concentration measure underlying NDS is not always well-suited to GMs with widely separated modes,
suitable modifications and extensions of the NDS test, e.g. to examine consistency of different local GM `submixtures', are also of interest. 
%


\printbibliography

\end{document}